\documentclass{aa}   
\usepackage{graphicx}

\newcommand{\beq}{\begin{equation}}
\newcommand{\eeq}{\end{equation}}
\newcommand{\beqn}{\begin{eqnarray}}
\newcommand{\eeqn}{\end{eqnarray}}
\newcommand{\lppr}{\stackrel{<}{\scriptstyle \sim}}
\newcommand{\gppr}{\stackrel{>}{\scriptstyle \sim}}

\usepackage{txfonts}
\begin{document}
   \title{Short-term VHE variability in blazars: PKS 2155-304}

   \author{F.M. Rieger
          \inst{1,2}
          \and
           F. Volpe\inst{1}}

   \offprints{F.M. Rieger}

   \institute{Max-Planck-Institut f\"ur Kernphysik, Saupfercheckweg 1, 69117 Heidelberg, Germany;
              \email{frank.rieger@mpi-hd.mpg.de}
         \and European Associated Laboratory for Gamma-Ray Astronomy, jointly supported by CNRS and MPG}

   \date{Received 02/2010; accepted 07/2010}

%flickering in the radio regime happens on timescales of days to weeks with amplitudes of a few percent
 
  \abstract
   {The $\gamma$-ray blazar PKS 2155-304 has attracted considerable attention because of its extreme TeV 
   variability characteristics during an exceptional flaring period in 2006. Among the observed key findings are 
   (i) a minimum variability timescale as short as $\sim 200$ sec and (ii) highly variable TeV emission, which in 
   the frequency interval [$10^{-4}$ Hz, $10^{-2}$ Hz] can be described by a log-normal distribution and 
   suggests an underlying multiplicative (and not additive) process.}  
  {Simultaneously accounting for these findings appears difficult within conventional approaches. Following earlier 
   suggestions for the TeV blazar Mkn 501 (Rieger \& Mannheim 2003), we explore a possible scenario where 
   PKS 2155-304 is supposed to harbor a supermassive binary black hole system and where the observed TeV 
   variability is dominated by emission from the less massive black hole.}
   {We analyze the constraints on the very high energy (VHE) source imposed by the observed variability characteristics 
   and the integrated VHE luminosity output, and discuss its implications for a binary black hole system.}
   {We show that for a secondary mass of $m_{\rm BH} \sim 10^7 M_{\odot}$, fluctuations in the disk accretion rate 
   that feed the jet could account for the observed red-noise type variability process down to frequencies of $\sim 10^{-2}$ 
   Hz. Jet curvature induced by orbital motion, on the other hand, could further relax constraints on the intrinsic jet speeds.}
   {Because a binary system can lead to different (yet not independent) periodicities in different energy bands, a longterm 
    (quasi-) periodicity analysis could offer important insights into the real nature of the central engine of PKS~2155-304.}
   

   \maketitle

\section{Introduction}
Active Galactic Nuclei (AGNs) show significant variability over a large range of timescales. Usually, a variety of 
information can be extracted from the data with statistical methods based on Fourier techniques. In the X-ray 
domain, for example, variability in accreting compact objects has commonly been described by means of power 
spectral densities (PSD), characterizing the amount of variability power $P(\omega)$ as a function of temporal 
frequencies $\omega$, or timescales $1/\omega$ (e.g., van der Klis 1997). For (Seyfert) AGNs the resultant PSDs 
quite often appear to follow power laws $P(\omega)\propto \omega^{-\beta}$, which on long timescales (small 
frequencies) are approximately described by $\beta$ close to $1$ (flicker noise), but break to a steeper slope 
($\geq 2$) on timescales shorter than a break timescale $t_B$. Active Galactic Nuclei thus vary more strongly 
towards lower frequencies (longer timescales). Recent studies have shown that the X-ray PSDs of AGNs can 
be qualitatively similar to the high state of black hole X-ray binary systems (e.g., McHardy et al. 2004, 2006). 
Based on these and other similarities, AGNs have sometimes been interpreted as scaled-up Galactic black 
hole systems.\\
In the very-high-energy (VHE) domain, the experimental situation is usually much less favorable. However, one
object where it became recently possible to employ similar analysis techniques is the TeV blazar PKS 2155-304 
($z=0.116$). Usually detected only with a low VHE flux of  $\sim 10\%$ of the Crab nebula, PKS 2155-304 
underwent a dramatic outburst in July 2006, with VHE flux levels varying between 1 and 15 Crab units, allowing 
an unprecedented variability analysis (Aharonian et al. 2007; Abramowski et al. 2010; Degrange et al. 2008). 
The Fourier analysis of its VHE light curve of, e.g., MJD 53944 indicates a red (Brownian) noise-type VHE PSD with 
an exponent close to 2 within the frequency range $[10^{-4}$ Hz, $10^{-2}$ Hz] (Aharonian et al. 2007). Similar 
results have been obtained in X-rays (1.5-10 keV) with BeppoSAX in 1996 and 1997 (Zhang et al. 1999). Moreover, 
similar to findings in the X-ray domain (Zhang et al. 1999, 2005), where linear relations between (absolute) rms 
variability amplitude and (mean) X-ray flux have been previously reported, a significant rms-flux correlation has 
been observed in the VHE data set (Degrange et al. 2008). This is known to be characteristic of a non-linear, 
log-normal stochastic process where the relevant, normally distributed variable is the logarithm of the flux 
$\log(X)$, and not just the flux itself (see Uttley et al. 2005 for a general treatment, and Superina et al.~2008 for 
application to PKS~2155). A log-normal distribution can be thought of as the result of many multiplicative random 
effects, whereas additive effects would give rise to a normal (Gaussian) distribution. Indeed, if the variability would 
be caused merely by an additive stochastic process, the fluxes would be normally distributed and no linear rms-flux 
relation would be expected, because all Fourier coefficients would be statistically independent. Therefore, the finding 
that the variations on short timescales (determining the rms) decrease in amplitude when the long timescale 
variations (determining the mean flux) decrease, indicates that the process driving the variations is a multiplicative 
process (as in a cascade) and not just an additive one (e.g., shot-noise), nor one resulting from independent 
variations in many separate regions (Uttley et al. 2005; McHardy 2008).\\
In accreting galactic black hole (BH) systems these variations have been frequently related to small, independent 
fluctuations in the accretion rate, occurring on local viscous timescale $t_{\rm visc}(r)$ over a range of disk 
radii $r$ that are large compared to the inner radius of the disk (Lyubarskii 1997; King et al. 2004; Arevalo \& Uttley 
2006). If not damped, these fluctuations can propagate inwards and couple together in a way to produce the 
multiplicative characteristics noted above. Any emission process linked to the the innermost region (e.g., as X-ray 
production or jet launching site) may then eventually be modulated over a frequency interval ranging from the 
inverse accretion time near the outer to the one at the innermost disk radius, respectively. While this scenario 
sounds very attractive (cf. also Giebels \& Degrange~2009 for BL Lac), a generalization to a supermassive 
source like PKS 2155-304 seems challenging given the detected VHE minimum variability (doubling) timescale 
of $t_v \sim 200$ sec. As we show below, a consistent approach along this line may require the presence of a 
putative binary BH system in PKS 2155-304 (cf. Rieger \& Mannheim 2003). It is this conjecture which we wish to 
explore here in more detail.

\section{The origin of variability in PKS 2155-304}
According to the VHE variability analysis for July 2006, there are two major findings which need
explanation: (i) First, the extreme short-term variability on a timescale as low as $t_v \sim 200$ sec (in 
the observer's frame)\footnote{As shown by, e.g., Degrange et al. (2008), this value is dependent on 
the telescope's sensitivity, which makes it possible to search for even faster variability with future arrays. 
The currently derived "doubling timescale" of $\sim 200$ sec is therefore best understood as an experimental 
upper limit on the real physical doubling timescale of the source.}, and (ii) secondly, the log-normal 
distribution of fluxes, which points to an underlying, multiplicative process. Consider the first finding (i):\\
If the luminosity of the host galaxy of PKS 2155-304 is indeed as high as reported, i.e., $M(R) \simeq 
-24.4$ mag (Kotilainen et al. 1998), its central BH mass is (even if one takes the scatter in the $M_{\rm 
BH}$ - $L_{\rm bulge}$ relation into account) expected to exceed (McLure \& Dunlop 2002)
\beq
M_t \geq 2  \times 10^8 M_{\odot}\,.
\eeq Accounting for the observed rapid variations based on conventional assumptions has thus proved 
challenging (see, e.g., Levinson~2007 and Begelman et al. 2008 for detailed discussions): 
In the galaxy's rest frame, the characteristic scale of any disturbance produced by the central engine 
cannot be smaller than its minimum size $r_g$, so that the minimum variability timescale for temporal 
fluctuations of the ejected fluid is $r_g/c$, where $r_g=GM/c^2$ is the gravitational radius 
(Levinson 2008). Only if parts of the jet flow would collide with a small external disturbance, e.g. a 
cloud, may smaller variability times get imprinted. Yet, in this case the bulk power that can be tapped for 
producing radiation will be correspondingly smaller, and the noted PSD shape and lognormal flux 
distribution may still be in need of explanation. If the observed minimum variability would indeed reflect 
the size of the central engine, $t_v \sim (1+z) r_g/c$, the allowed maximum BH mass would be almost 
one order of magnitude smaller than the one inferred from the host galaxy luminosity relation, i.e., 
\beq\label{m_var}
M_v  \lppr 4 \times 10^7 \left(\frac{t_v}{200~\mathrm{sec}}\right)~M_{\odot}\,.
\eeq Similar results are obtained if the variability is taken to be produced by the collision of shells moving 
at different speeds (as in classical internal shock scenarios). In all these cases, the inferred minimum 
source size (in the rest frame of the galaxy) becomes independent of any Doppler or Lorentz factor.\footnote{Note 
that the situation could be different if the variations are produced within the relativistically moving component 
itself, e.g., via a cascade induced by internal $\gamma\gamma$-absorption, cf. Aharonian et al. (2008).} 
Interestingly, similar small BH masses $M_v \lppr 3 \times 10^7 M_{\odot}$ for PKS 2155-304 have 
been previously inferred based on its X-ray variability properties (Hayashida et al. 1998; Czerny et al. 2001). 
Moreover, Lachowicz et al. (2009) have recently suggested that the detection of a possible $\sim4.6$ hr 
X-ray quasi-periodic oscillation (QPO) with XMM-Newton in 2006 might be related to a hot spot orbiting in 
the disk close to the innermost stable orbit. If this interpretation is correct, is would imply a firm upper 
limit on the BH mass of $\leq 3 \times 10^7 M_{\odot}$ (Schwarzschild) and $\leq 2.1\times 10^8 M_{\odot}$ 
(extreme Kerr), respectively.\\ 
One obvious possibility to reconcile these apparently divergent central mass estimates is related to the 
putative presence of a close binary BH system where the jet that dominates the high-energy emission originates 
from the less massive (secondary) BH. In such a case, host galaxy observations may only be indicative of the 
total (primary and secondary) BH mass, whereas high-energy observations could actually reveal signatures 
of the less massive BH. We note  that the possible presence of close binary BH systems in blazar-type AGN 
sources has been repeatedly invoked to account for a variety of observational findings, most notably for the 
detection of mid-and long-term quasi-periodic variability (e.g., see Rieger~2007 for a review).\\ 
If the observed VHE output in PKS 2155-305 were be dominated by emission from the secondary BH, 
however, its mass cannot be supposed to be too small if one also wishes to account for the required jet 
power. Suppose that the maximum possible jet power is constrained by the Eddington accretion limit $L_E 
\sim 10^{46} (M_{\rm BH}/10^8 M_{\odot})$ erg/s. The maximum spin power, for example, which can be 
extracted from the rotational energy of a BH (via a Blandford-Znajek-type process) is on the order of $L_s 
\sim 10^{44} (M_{\rm BH}/10^8 M_{\odot})^2 (B_H/10^4 \mathrm{G})^2$ erg/s. With an Eddington equipartition 
magnetic field strength $B_H \propto 1/\sqrt{M_{\rm BH}}$, the extractable spin power thus becomes 
comparable to $L_E$, i.e., $L_s \sim L_E/3$. The average VHE gamma-ray flux detected by H.E.S.S. implies 
an isotropic VHE luminosity $L_{\rm iso} \sim (5-10) \times 10^{46}$ erg/s (Aharonian et al. 2007). If the 
radiation is produced within two, oppositely directed, conical jets of an opening angle $\theta \lppr 0.1$ rad, 
the required (minimum) jet power would be $L_j \sim (1-2)\times10^{45} (\theta/0.1)^2$ erg/s. For a BH 
accreting at close to the Eddington limit (see below) or with maximal spin power, the required minimum 
mass thus becomes
\beq\label{m_l}
  M_{\rm l} \sim 10^7~M_{\odot} \left(\frac{\theta}{0.1}\right)^2\,, 
 \eeq which would be consistent with the variability constraints (Eq.~\ref{m_var}) and be suggestive of a 
 binary mass ratio $m_2/m_1 \simeq M_{\rm BH}/M_t > 0.01$.\\ 
 In reality, the variability constraint of Eq.~(\ref{m_var}) is probably too optimistic, in particular if one wishes to 
 account for the second finding (ii) (i.e., the log-normal VHE flux distribution) based on fluctuations in the 
 accretion flow that feeds the jet. Consider again the fluctuating disk model (Lyubarskii 1997), where fluctuations 
 of the disk parameters at some radius, which occur on local viscous timescale, can lead to variations in the 
 accretion rate at smaller radii that are of the flicker- or red-noise type. Then the relevant timescale 
 to be employed is the viscous timescale $t_{\rm visc}$ close to the inner radius of the disk, and not the 
 dynamical timescale $r_g/c$. In terms of the $\alpha$ parameter and the disk scale height to radius ratio 
 $(h/r)$, this timescale can be expressed as
\beq
 t_{\rm visc} = \frac{1}{\alpha}\left(\frac{r}{h}\right)^2 \left(\frac{r}{r_g}\right)^{3/2} \frac{r_g}{c}\,,
 \eeq where $\alpha$ is the viscosity coefficient. Consequently, the inner disk needs to be sufficiently thick 
and viscous ($(h/r)^2\alpha \gppr 0.1$) in order to minimize the variability timescale without violating the 
luminosity constraint eq.~(\ref{m_l}). Similar conditions seem to be required to account for the X-ray 
spectral-timing properties of both AGNs and BH X-ray binaries (Arevalo \& Uttley 2006). Suppose therefore 
that in its innermost parts the disk is geometrically thick (as, e.g., for an ADAF). Within a more global scenario, 
one may then naturally expect a transition to a standard cooling-dominated (geometrically-thin) disk to occur 
at some radius $r_{\rm tr}$. These transition regions are known to be prone to dynamical instabilities (e.g., 
Gracia et al. 2003). Interestingly, on longer timescales the X-ray (RXTE) PSD of PKS~2155-304 seems to show 
a characteristic rollover-timescale of $t_c \sim 1$ day (Kataoka et al.~2001). If this is indicative of the location 
of the transition region, i.e., $t_{\rm visc}(r_{\rm tr}) \sim t_c$, a reasonable transition radius $r_{\rm tr} \sim 
50~r_g$ might be inferred (note that for ADAFs $(h/r) \sim 1$ throughout).\\ 
In order to account for the observed VHE variability characteristics, suppose therefore that fluctuations in the 
disk accretion rate on timescales as short as $t_v$ are efficiently transmitted to the jet, leading to red noise-type 
fluctuations in the injection rate for Fermi-type particle acceleration. Obviously we will only be able to observe 
emission with red noise-characteristics if these signatures do not get blurred by processes occurring on a 
longer timescale within the source: For an observer, flux changes will always appear to be convolved and 
thus dominated by the longest timescales (Salvati et al. 1998), i.e.,
\beq
\Delta t_{\rm obs} \simeq \mathrm{max}\{t_v, \Delta t_r, \Delta t_{\rm rad}\}\,,
\eeq where $t_v$, $\Delta t_r$ and $\Delta t_{\rm rad} \geq t_{\rm acc}$ are the timescales (as seen by an 
observer) for injection, for photons traveling across the radial width of the source, and for the relevant radiative 
processes, respectively. For the radial travel time across the source one has
\beq
 \Delta t_r \simeq \frac{r}{\Gamma_b c}\,, 
 \eeq where $\Gamma_b \geq 10$ denotes the characteristic bulk Lorentz factor and $r$ the radial source 
 dimension in the rest frame of the galaxy (note that length contraction only affects the dimension along the 
 direction of motion). Typical bulk Lorentz factors inferred for the VHE emission region(s) in PKS 2155-305 
 during the 2006 high state are in the range of $\Gamma_b \sim (30-50)$ and perhaps even higher (e.g., Begelman 
 et al.  2008). For the model to be consistent, one thus requires a radial source dimension that satisfies $r < 2 \times 
 10^{14}~(t_v/200~\mathrm{sec}) (\Gamma_b/30)$ cm. Either the VHE region is thus located very close to the BH, 
 or the jet exhibits an internal velocity stratification of the (fast) spine - (slow) shear-type where the VHE emission 
 is e.g. produced by Compton up-scattering within a fast moving spine, similar to the needle-in-jet model 
 recently proposed (Ghisellini \& Tavecchio 2008). \\
 Consider next the radiative timescale and assume synchrotron losses to dominate the high-energy branch of 
 the energetic particle distribution $n_e(\gamma')$ that is responsible for the up-scattering of soft photons to the 
 VHE regime. The radiative timescale thus is 
 \beq
   \Delta t_{\rm rad} \simeq 40 \left(\frac{5 \times 10^5}{\gamma'}\right) \left(\frac{1~\mathrm{G}}{B'}\right)^2
                                                \left(\frac{40}{\Gamma_b}\right)~\mathrm{sec}\,.
 \eeq Hence, in order to work successfully, the proposed scenario requires co-moving magnetic field strengths that 
 are sufficiently high ($B' \sim 1$ G) and electrons responsible for up-scattering that are sufficiently energetic ($\gamma' 
 \gppr 10^5$). These parameters are again consistent with recent model fits to the observed spectral energy distribution 
 of PKS 2155-304 within the needle-in-jet framework (Ghisellini \& Tavecchio 2008). Obviously, extreme short-term 
 variability may for these (and other) reasons not necessarily show up at lower energies. As a consequence: 
 If synchrotron radiation from the fast "needle" component would dominate the emission in the optical domain, then 
 the detectable minimum variability timescales are expected to be correspondingly longer, e.g., on the order of a few 
 to several tens of minutes for the above noted parameters. This might be compared with the minimum variability 
 timescale of $\sim 15$ min during an optical observational campaign on PKS 2155-304 in 1995 (Paltani et  al. 1997).

\section{Possible implications of a binary BH model}
\begin{figure}
\centering
\includegraphics[width=8cm]{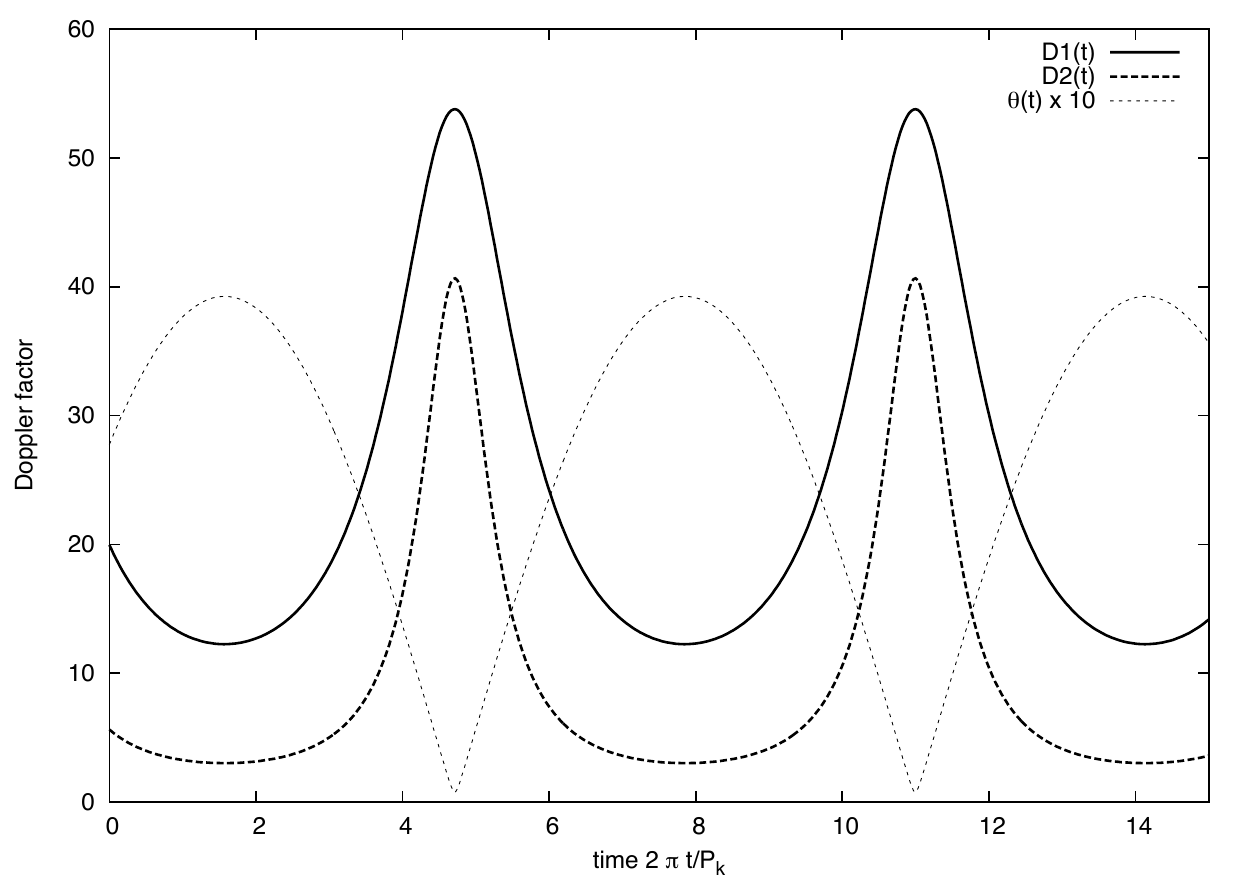}
\caption{Time-dependence of the Doppler factor for motion along an idealized orbital-driven helical jet path. Chosen 
parameters are: $\Gamma_{b,z}=20$, $i=2^{\circ}$, $P_k \simeq 13$ yr (thick solid line), and $\Gamma_{b,z}=20$, 
$i=4^{\circ}$, $P_k \simeq 4$ yr (thick dashed line). Also shown is the time-dependence of the angle $\theta(t)$, the
actual angle of the emission region to the line of sight (in degree, multiplied by a factor 10) for $\Gamma_{b,z}=20$,
$i=2^{\circ}$ and $P_k \simeq 13$ yr.}\label{fig2}
\end{figure}

Following the above analysis, let us suppose that PKS 2155-304 harbors a close binary system consisting of a 
primary BH with mass $m_1 \sim 5 \times 10^8 M_{\odot}$ and a secondary one with mass $m_2\sim10^7 M_{\odot}$ 
(mass ratio $q \sim0.02$). Evidence for a possible longterm period of $\sim(4-7)$ yr in the optical V band (Fan \& 
Lin 2000; cf. also Brinkmann et al. 2000 for soft-X-ray hints) might possibly fit well into such a framework: 
If one requires the secondary to be on an orbit that could intersect the accretion disk around the primary with a typical 
(maximum) disk size of $10^3 x~r_s$, with $x \sim 1$ and $r_s=2~G m_1/c^2$ (Goodman 2003), possible Keplerian 
periods $P_k=2\pi/\Omega_k$ are constrained to be smaller than $P_k \simeq 45 \,(1+q)^{-1/2} x^{3/2} (m_1/5 \times 
10^8 M_{\odot})$ yr, a condition that seems well satisfied for PKS 2155-304. This suggests that binary disk interactions 
could indeed result in optical QPOs on the timescale of several years (see Rieger 2007 for more details). Accordingly, 
we may derive an upper limit on the intrinsic Keplerian orbital period of the binary 
\beq\label{kepler}
 P_k \leq  \frac{2}{(1+z)}\,P_{\rm obs}^{\rm opt} \simeq 13 \left(\frac{P_{\rm obs}^{\rm opt}}{7~\mathrm{yr}}\right)~\mathrm{yr}\,,
\eeq by assuming that the observed optical longterm periodicity is caused by the secondary crossing the disk around 
the primary twice per orbital period, a situation expected to occur during the initial stage of interaction. In reality, 
however, this may not be fully appropriate because an alignment of the binary's orbital and primary's disk plane typically 
occurs on comparatively short timescales (e.g., Ivanov et al. 1999). One may thus expect quasi-coplanar orbits, with the 
binary possibly surrounded by a circumbinary disk, to be much more common. Yet, granted that to be the case, periodically 
modulated mass transfer across the gap (see below) could still lead to QPOs in the standard disk-dominated 
infrared-optical regime. Most circumbinary disk simulations suggest that the accretion rate is pulsed over one orbital 
period, although occasionally two peaks per period are found (e.g., Artymowicz \& Lubow 1996; Bate \& Bonnell~1997; 
Kley~1999; Hayasaki et al. 2007). This indicates that Eq.~(\ref{kepler}) still provides us with a reasonable constraint on 
the maximum intrinsic orbital period. If so, then the residual lifetime $t_g$ of the system due to emission of gravitational 
waves (assuming a quasi-circular orbit) would be in the range $t_g \propto P_k^{8/3} \sim (3 \times 10^6 - 8 \times 10^7)$ 
yr. The current separation of the putative binary, on the other hand, would be on the 
order of
\beq 
   d \simeq 6 \times 10^{16} \left(\frac{P_k}{13\,\mathrm{yr}}\right)^{2/3} \left(\frac{m_1}{5 \times 10^8\,M_{\odot}}\right)^{1/3}
                    \mathrm{cm}  \sim (200-400)~r_s \,.
\eeq Once the secondary becomes embedded in the outer disk around the primary, it starts clearing up an annular gap of 
a radial size $\Delta r \sim (h/d)\,d\,\alpha^{-0.4}$ in the disk as soon as the timescale for gap-opening via tidal torques 
becomes smaller than the timescale on which diffusion can refill it (Papaloizou \& Lin~1984). It was initially thought that 
this would cut off the binary BHs from any mass supply (accretion) of the circumbinary disk. However, numerical simulations 
show that even with significant gap, mass can be supplied to the central binary through tidal, time-dependent gas streams 
that penetrate the disk gap, periodically approaching and preferentially feeding the secondary (see references above). 
Accordingly, for non-vanishing binary eccentricity, QPOs with period equal to the orbital one may arise due to a 
time-modulated accretion rate. The simulations indicate in particular that a reversal of mass accretion rates can occur (i.e., 
$\dot{m_2} > \dot{m_1}$ despite $m_2 <m_1$), acting towards equal-mass binaries and suggesting that the secondary may 
become more luminous than the primary. If that is the case, then applying variability constraints based on observed emission properties can lead to a spurious identification of the secondary mass with the (total) central BH mass. As suggested above, 
this may help to explain some of the discrepancies in black mass estimates based on high-energy emission models and 
host-galaxy observations.\\
Suppose that the size $r_d$ of the disk around the secondary is comparable to the Roche lobe radius (Eggleton~1983) 
or the Toomre stability radius (whichever is smaller). For $m_2 \sim 10^7 M_{\odot}$ and $r_d \sim 10^3 r_g$, the viscous 
timescale at the outer radius $t_{\rm visc}(r_d) \sim 50~(0.1/\alpha) ([r/h]/10)^2$~yr (standard disk) then most likely 
exceeds the orbital period $P_k \sim10$ yr. Any modulation of the mass supply to the outer disk at the orbital period may 
thus be damped by the viscosity of the disk, i.e., as the next pulse comes before the former was able to propagate in, the 
amplitude of variability becomes small as a result of summation of the pulses (Zdziarski et al. 2009). A significant modulation 
of the jet power (emerging from the innermost part of the disk) solely owing to a periodically changing mass supply (at the 
outer disk radius) may therefore not become apparent. On the other hand, if variability in the disk occurs on thermal 
timescale $t_{\rm th} \sim \alpha^{-1} (r/r_g)^{3/2} r_g/c$, optical (V band) disk variability on orbital timescale might be 
possible. While a periodically changing mass supply may thus not necessarily lead to significant periodicities in the jet 
emission, geometrical effects due to the orbital motion could introduce them: If the jet launched is wrapped by a sufficiently 
strong magnetic field, for example, its overall path can become curved due to the orbital motion of the jet-emitting BH 
(leading to non-ballistic-type helical motion). Then the constraints on the required minimum bulk Lorentz factors of 
the outflow ($\Gamma_{b,z}$, perpendicular to the disk plane) might be somewhat relaxed because the effective Doppler 
factor will be time-dependent, $D(t)$, with the strongest boosting effects occurring whenever the (changing) angle to the line 
of sight becomes smallest (cf. Rieger 2004). This is illustrated in Fig.~\ref{fig2} for a bulk Lorentz factor (in the rest frame of the 
secondary) of $20$ and an inclination $i$ (defined as the angle between the line of sight and the z-axis, in a frame where the 
orbital plane is in the x-y plane) between $2^{\circ}$ and $4^{\circ}$. Obviously, considerable changes in the effective Doppler 
factor could occur, which would further influence the detectability of high- and low-level (intrinsic) active source stages, i.e., 
even if the intrinsic flux would change only moderately, an order of magnitude difference in measured VHE fluxes may 
become possible. 
Moreover, the motion of components along curved trajectories could possibly also account for the observed super-quadratic 
correlation between the X-ray and VHE flux (Costamante~2008; Katarzynski \& Walczewska 2010). Of course, this might only 
be expected as long as the lateral width of the jet remains smaller than the separation $d$ of the binary, so that the effects 
caused by orbital motion will not be smeared out. For an expanding jet, this obviously limits the distance up to which extreme 
Doppler effects may occur. This might partly explain why on larger (radio VLBA) scales, only modest Doppler boosting 
seems to be present (e.g., Piner \& Edwards 2004). Clearly, an accurate identification of active VHE source stages in PKS 
2155-304 could help to further assess the influence of differential (geometrical) Doppler effects.   

\section{Conclusions}
The observed, extreme VHE variability characteristics of PKS~2155-304 provide strong constraints on the physical 
parameters of its engine. We have suggested that similar to Mkn 501 (Rieger \& Mannheim 2003), the putative 
presence of a close supermassive binary BH system could allow (i) to reconcile central mass estimates based on 
host galaxy observations (indicative of the total primary and secondary mass) with those based on VHE 
$\gamma$-ray variability (possibly only indicative of the jet-emitting secondary), (ii) to account for the observed 
log-normal variability characteristics via accretion disk fluctuations, and (iii) to relax constraints on the jet flow velocity. 
Obviously, an increased instrumental sensitivity in the TeV domain (by, e.g., a CTA-type instrument) that may allow us
to search for even faster variability and, complementary, an advanced QPO analysis in the optical (e.g., Hudec \& 
Basta 2008) could thus be particularly useful to further assess the possibility of a binary scenario for PKS 2155-304. If 
corroborated by further observations, the analogy to X-ray binary systems might be closer than initially anticipated.

\begin{acknowledgements}
Discussions with Felix Aharonian, Amir Levinson and Berrie Giebels are gratefully acknowledged.   
\end{acknowledgements}

\end{document}